\RequirePackage{fix-cm}
\documentclass[smallextended]{svjour3}       
\smartqed  
\usepackage{graphicx}
\usepackage{mathptmx}      
\usepackage{latexsym}
\usepackage{amsfonts}
\usepackage{latexsym}
\usepackage{amsmath}
\usepackage{amssymb}
\journalname{...}
\begin{document}

\title{Dispersive quantum systems
}
\subtitle{A class of isolated non-time reversal invariant quantum systems}


\author{L\'{u}cio Fassarella
}

\institute{Universidade Federal do Esp\'{i}rito Santo \at
              CEUNES - Rodovia BR 101 Norte, Km. 60, CEP 29932-540, S\~{a}o Mateus -- ES, Brazil \\
              Tel.: +55(27)3312-1511\\
              Fax: +55(27)3312-1510\\
              \email{lucio.fassarella@ufes.br}           
}

\date{September-01/2011}

\maketitle

\begin{abstract}
A \textit{dispersive quantum system} is a quantum system which is both
isolated and non-time reversal invariant. This article presents precise definitions for
those concepts and also a characterization of dispersive quantum systems
within the class of completely positive Markovian quantum systems in finite
dimension (through a homogeneous linear equation for the non-Hamiltonian
part of the system's Liouvillian). To set the framework, the basic features of quantum mechanics are reviewed focusing on time evolution and also on the theory of completely positive Markovian quantum systems, including
Kossakowski-Lindblad's standard form for Liouvillians. After those general
considerations, I present a simple example of dispersive two-level
quantum system and apply that to describe neutrino oscillation.
\keywords{quantum time evolution \and non-time reversal invariance \and isolated quantum system \and neutrino oscillation}
\PACS{03.65.Aa  \and 03.65.Yz \and 13.15.+g}
\end{abstract}

\section{Introduction}
\label{intro}
The search for a scientific understanding of \textit{time} has a wide scope
in physics, ranging from classical mechanics to quantum field theory, from
particle mechanics to cosmology, statistical physics and beyond. The
research on time inevitably touches foundational issues and one can even
suspect it cannot be fully understood since time is so essential to our
perception of reality. Nevertheless, we can hope to improve our knowledge
about time as \textit{time goes on}...

Instead of to deal with the subtleties of the physical meaning of \textit{time} (for detailed discussion see \cite{Rov} and \cite{Ze} and references quoted therein), here I'm devoted to a simpler task: to show that \textit{it is theoretically possible an elementary quantum system be both isolated and non-time reversal invariant}. This possibility contradicts a common sense among physicists, namely, that isolated systems are (ever) time reversal invariant and that irreversibility is just a statistical phenomenon (coded in the Second Law of
Thermodynamics).

\medskip

To be more precise, consider a quantum system and denote its state (density
operator) in time $t$ by $\rho \left( t\right) $ -- here, I use
\textit{Schr\"{o}dinger's picture}. It is generally accepted that if the system
is \textit{closed}, then its time evolution is given by von Neumann's
equation with some\ time-dependent\ Hamiltonian $H\left( t\right) $:%
\begin{equation}
\frac{d}{dt}\rho \left( t\right) =-\frac{i}{\hbar }\left[ H\left( t\right)
,\rho \left( t\right) \right]  \label{equation_von-neumann}
\end{equation}%
Accordingly, the system is said to be \textit{isolated} when it is closed and its
Hamiltonian is constant. So, according with this view \textit{open quantum systems} are exactly those whose time evolution is not governed by (time-dependent) von Neumann equation (\ref{equation_von-neumann}). In books and papers those conceptions may be used implicitly; for
example, authors can relate the semigroup structure of time evolution to (Markovian) open
systems \textit{only}.\footnote{%
There are many examples, but I mention only the book by Breuer-Petruccione 
\cite[p.110]{BP}, the article by G.E. Crooks \cite{Cr} and some pioneers on the subject: V. Gorini at al. \cite{GKS}, G. Lindblad \cite{Li},  Mehra-Sudarshan \cite{Me}, Kossakowski \cite{Kos} and Ingarden-Kossakowski \cite{IKos}.}

Although the above definitions for the concepts of \textit{isolated}, \textit{closed} and \textit{open} can be mathematically perfect, they lack direct correspondence with those physical meanings we give to them. Indeed, \textit{when we say a
physical system is closed we mean that it does not exchange matter with other systems}; also, \textit{we say a physical system is isolated when it does not interact with any other system}. In this sense, one realizes that the concept of isolated system must be related to the Principle of Inertia, meaning that \textit{after arbitrary preparation, an isolated system has its energy and momenta remaining constant}. Assuming the operationalist point of view, I define the concept of isolated quantum systems in terms of the expectation values of its energy-momentum tensor operator:

\begin{definition}
A quantum system is isolated when the expectation values of its energy-momentum tensor operator with respect to any time-dependent state is constant.
\end{definition}

I argue that this definition is not trivially irrelevant, even though it is a truism that there is no isolated system (within Universe) -- at least, they could not be observed even if they existed.\footnote{According to Breuer-Petrucione \cite[p.vii]{BP}: \textquotedblleft Quantum
mechanical-systems must be regarded as open systems. On the one hand, this
is due to the fact that, like in classical physics, any realistic system is
subjected to a coupling to an uncontrollable environment which influences it
in a non-negligible way. The theory of open quantum systems thus plays a
major role in many applications of quantum physics since perfect isolation
of quantum systems is not possible and since a complete microscopic
description or control of the environmental degrees of freedom is not
feasible or only partially so.\textquotedblright} Nevertheless, many systems can
be regarded isolated in practice, at least for a short
interval of time, and this is sufficient to the concept be (eventually) useful.

Beyond the almost triviality of give a definition for the concept of isolated quantum system, I'm going to highlight a special class of them, namely, \textit{isolated quantum systems that are non-time reversal invariant}.\footnote{\textit{Non-time reversal invariance} is defined in section \ref{sec:1}.} Quantum systems of this sort I call \textit{dispersive quantum systems}:
\begin{definition}
A dispersive quantum system is a quantum system which is isolated and non-time reversal invariant.
\end{definition}
Perhaps, a distinguished example of dispersive quantum system
is the whole Universe.\footnote{Actually, in order to the Universe be consistently regarded isolated we must take gravity into account, as it was noted by Landau and Lifschitz in \cite[p.30]{LLs}; otherwise, we should describe matter as under the influence of ever changing external conditions.} However, it would be remarkable if we could
discover a dispersive quantum system being also elementary,
because its non-time reversal invariance would be fundamental -- in the sense of not
being an emergent (statistical) property. The \textit{dispersive qubit} (section \ref{sec:2}) and its application to \textit{neutrino oscillation} (section \ref{sec:3}) are intended to instantiate that!

\medskip

The structure of the paper is simple. Section \ref{sec:1} starts with general
quantum mechanics focusing on time evolution and a definition of
irreversibility, followed by a review of definitions and results about
Markovian quantum systems, dynamical semigroups and completely positiveness.
After, it is obtained the equation for the Liouvillian's non-Hamiltonian
part that characterizes those completely positive Markovian quantum systems
that are isolated. In section \ref{sec:2}, it is presented an explicit model
exhibiting the mentioned features. In section  \ref{sec:3} I apply previous developments to describe neutrino oscillation, with the introduction of a new parameter I call \textit{dispersive parameter}. In the final section \ref{sec:4}, some remarks are discussed.

\medskip

\begin{remark}
I denote by $\mathbb{N}$ the set of natural numbers including zero and set $%
\mathbb{N}^{\ast }:=\mathbb{N}\backslash \left\{ 0\right\} $. For a Hilbert
space $\mathcal{H}$, I denote by $\mathcal{L}\left( \mathcal{H}\right) $ the
set of densely defined operators in $\mathcal{H}$, by $\mathcal{B}\left( 
\mathcal{H}\right) $ the space of bounded operators in $\mathcal{H}$ and by $%
\mathcal{T}\left( \mathcal{H}\right) $ the space of bounded trace class
operators in $\mathcal{H}$. Finally, I use \textit{natural units}, so $\hbar
=1$.
\end{remark}

\section{Quantum mechanical systems}
\label{sec:1}
\subsection{Basic structure}
\label{sec:1.2}
In Quantum Mechanics, \textit{physical systems}\ are described in terms of 
\textit{observables} and \textit{states} with the use of a \textit{separable
Hilbert space} $\mathcal{H}$: for a system $\mathfrak{S}$ without \textit{%
superselection sectors}, observables are identified with \textit{(densely
defined) self-adjoint operators} in $\mathcal{H}$ and states are identified
with \textit{density operators}, \textit{i.e.}, positive trace class\textit{%
\ }operators with trace one in $\mathcal{H}$. I denote the set of
observables by $\mathcal{A}\left( \mathcal{H}\right) $ and the set of states by $\mathcal{S}\left( \mathcal{H}\right) $. 
The fundamental postulate of Quantum Mechanics states that %
\textit{the expectation value of an observable} $A\in \mathcal{A}\left( \mathcal{H}\right)$ %
\textit{\ when the system is in the state} %
$\rho \in \mathcal{S}\left( \mathcal{H}\right)$%
\textit{\ is given by}%
\footnote{There are some technical requirements to this formula to be well defined in
general, but it is always well defined when $A$ is bounded -- what is
automatic in the finite dimensional case, which is the one we are interested.}
\begin{equation}
\left\langle A\mid \rho \right\rangle :=tr\left( \rho A\right)
\end{equation}

\paragraph{Time evolution.}

\medskip

Consider the system was prepared at time $t_{0}$ and evolves without
interference since then. In Schr\"{o}dinger's picture, system's \textit{time
evolution} from an instant $t_{1}\geq t_{0}$ to a later instant $t_{2}$ must
be described by a map from the space of states to itself,%
\begin{equation}
\Gamma _{t_{2},t_{1}}:\mathcal{S}\left( \mathcal{H}\right) \rightarrow 
\mathcal{S}\left( \mathcal{H}\right) \ \ \left( t_{2}>t_{1}\right)
\label{def_time-evolution-map}
\end{equation}%
I call this \textit{time evolution map} and its physical interpretation is
simple:\textit{\ given two instants }$t_{2}>t_{1}\geq t_{0}$\textit{, if }$%
\rho_{1} $\textit{\ is the system's state at instant }$t_{1}$\textit{, then }$%
\Gamma _{t_{2},t_{1}}\rho_{1} $\textit{\ is the system's state at the instant }$%
t_{2}>t_{1}$. It is natural to assume the following property I call \textit{%
factorization}: 
\begin{equation}
\Gamma _{t_{3},t_{1}}=\Gamma _{t_{3},t_{2}}\Gamma _{t_{2},t_{1}}\ \ ,\
\forall t_{3}>t_{2}>t_{1}\geq t_{0}
\label{form_time-evolution_factorization}
\end{equation}

A physical system is \textit{time reversal invariant}\footnote{%
\textit{Time reversal invariance} is a property of systems, while \textit{reversibility} is a property of a system's states. In general, the non-time reversal invariance of a system is related to the existence of an irreversible state. See \cite{Jo} and \cite{ACL} for a more detailed discussion on the concepts of time reversal invariance and reversibility, as well as the
relation between them and the Second Law of Thermodynamics.} when there is a map $\Upsilon :\mathcal{S}\left( \mathcal{H}\right)
\rightarrow \mathcal{S}\left( \mathcal{H}\right) $ satisfying\newline
i) Idempotence:%
\begin{equation}
\Upsilon ^{2}=id  \label{def_reversible_idempotence}
\end{equation}%
ii)\ Time-reversing equation:%
\begin{equation}
\Gamma _{t_{2},t_{1}}\Upsilon \Gamma _{t_{2},t_{1}}=\Upsilon \ \ ,\ \forall
t_{2}>t_{1}\geq t_{0}  \label{def_reversible_time-reversing-equation}
\end{equation}%
I call $\Upsilon $ the \textit{time reversing map}. By physical reasons, one may require additional properties on $\Upsilon $, such as antilinearity. Its physical
meaning is natural: $\Upsilon $ defines a correspondence among the system's
states which reverts the direction of time evolution:
\begin{equation}
\Upsilon\Gamma _{t_{2},t_{1}}\Upsilon= \Gamma _{t_{2},t_{1}}^{-1} \ \ ,\ \forall
t_{2}>t_{1}\geq t_{0}
\end{equation}%
This equation is the quantum
analog of what can in principle be done to a classical (non-magnetic) system: if the
velocities of all particles of a classical mechanical system are reversed
(what can be represented by a map in phase space), then this system would
behave \textit{as if} it was running backwards in time!\footnote{%
In general, isolated classical systems are time reversal invariant; but for a system with
very large number of degrees of freedom, its macroscopic behavior exhibits
statistical properties that allow us to distinguish the past from the future.%
}

A system is \textit{non-time reversal invariant} when it is not time reversal invariant.
Time reversal invariance holds for some physical systems, but one does not have any
reason to assume this property must hold for all closed physical systems.
(Actually, I'm going to describe a class of \textit{isolated non-time reversal invariant quantum
systems.})

\medskip

If the system is time reversal invariant, equations (\ref{def_reversible_idempotence})
and (\ref{def_reversible_time-reversing-equation}) imply $\Gamma
_{t_{2},t_{1}}$ is invertible, for all $t_{2}>t_{1}\geq t_{0}$ -- so, 
\textit{time-evolution map's invertibility is a necessary condition for the
system to be invertible}:%
\begin{equation}
\Gamma _{t_{2},t_{1}}^{-1}=\Upsilon \Gamma _{t_{2},t_{1}}\Upsilon \ \ ,\
\forall t_{2}>t_{1}\geq t_{0}  \label{def_reversible_inversion}
\end{equation}%
In this case, one can define the \textit{extended time evolution map} in $%
\mathcal{S}\left( \mathcal{H}\right) $:%
\begin{equation}
\tilde{\Gamma}_{t_{2},t_{1}}:\mathcal{S}\left( \mathcal{H}\right)
\rightarrow \mathcal{S}\left( \mathcal{H}\right) \ \ ,\ \ \tilde{\Gamma}%
_{t_{2},t_{1}}:=\left\{ 
\begin{array}{r}
\Gamma _{t_{2},t_{1}}\ \ ,\ t_{2}>t_{1}\geq t_{0} \\ 
id\ \ \ \ ,\ \ t_{2}=t_{1}\geq t_{0} \\ 
\Gamma _{t_{2},t_{1}}^{-1}\ \ ,\ t_{1}>t_{2}\geq t_{0}%
\end{array}%
\right. \ \ ,\ \forall t_{1},t_{2}\geq t_{0}
\label{def_time-evolution-extended}
\end{equation}%
The extended time evolution map satisfies \textit{extended factorization}:%
\begin{equation}
\tilde{\Gamma}_{t_{3},t_{1}}=\tilde{\Gamma}_{t_{3},t_{2}}\tilde{\Gamma}%
_{t_{2},t_{1}}\ \ ,\ \forall t_{3},t_{2},t_{1}\geq t_{0}
\label{form_time-evolution-extended_factorization}
\end{equation}%
Proof: I have to analyze the six possible orderings for instants $%
t_{1},t_{2},t_{3}\geq t_{0}$; the case $t_{3}>t_{2}>t_{1}\geq t_{0}$ follows
directly from factorization (\ref{form_time-evolution_factorization}); here,
I verify explicitly the case $t_{2}>t_{3}>t_{1}\geq t_{0}$ only: 
\begin{equation*}
\tilde{\Gamma}_{t_{3},t_{1}}=\Gamma _{t_{1}t_{3}}=\Gamma
_{t_{2},t_{3}}^{-1}\Gamma _{t_{2},t_{3}}\Gamma _{t_{3},t_{1}}\ =\Gamma
_{t_{2},t_{3}}^{-1}\Gamma _{t_{2},t_{1}}\ =\tilde{\Gamma}_{t_{3},t_{2},}%
\tilde{\Gamma}_{t_{2},t_{1}}
\end{equation*}

\begin{remark}
For a time reversal invariant system, there is no intrinsic distinction between past and
future, since any pair of states related by time evolution map are equally
related by the time evolution map's inverse. However, an observer assigns a
\textquotedblleft time arrow\textquotedblright\ to an non-time reversal invariant system
through preparation time $t_{0}$: before $t_{0}$ the system interacts with
environment and after $t_{0}$ the system evolves without external
interaction \cite[p.32]{LLs}, \cite[\S 7]{LLq}.
\end{remark}

\medskip

\paragraph{Markovian systems.}

\medskip

A system is said to be \textit{Markovian} when its time evolution map
depends of the time interval between instants only:%
\begin{equation}
\Gamma _{t_{2},t_{1}}=\Gamma _{t_{0}+t_{2}-t_{1},t_{0}}\ \ ,\ \forall
t_{2}>t_{1}\geq t_{0}  \label{def_markovian-property}
\end{equation}%
In this case, the system's time evolution is given by the \textit{%
(one-parameter) quantum dynamical semigroup} in $\mathcal{S}\left( \mathcal{H%
}\right) $ with domain $\left[ 0,\infty \right) $:%
\begin{equation}
\Gamma _{t}:\mathcal{S}\left( \mathcal{H}\right) \rightarrow \mathcal{S}%
\left( \mathcal{H}\right) \ \ ,\ \Gamma _{t}:=\left\{ 
\begin{array}{r}
id\ \ \ \ \ \ \ \ ,\ t=0 \\ 
\Gamma _{t_{0}+t,t_{0}}\ ,\ t>0%
\end{array}%
\right.  \label{def_quantum-dynamical-semigroup}
\end{equation}%
Directly from factorization (\ref{form_time-evolution_factorization}), it
follows the \textit{semigroup property}:
\begin{equation}
\Gamma _{t_{2}}\Gamma _{t_{1}}=\Gamma _{t_{2}+t_{1}}\ \ ,\ \forall
t_{1},t_{2}\geq 0 \label{form_time-evolution-semigroup_factorization}
\end{equation}%
Proof: $\Gamma _{t_{2}}\Gamma _{t_{1}}=\Gamma _{t_{2}+t_{1},t_{1}}\Gamma
_{t_{1},0}=\Gamma _{t_{2}+t_{1},0}=\Gamma _{t_{2}+t_{1}}\ \ ,\ \forall
t_{1},t_{2}>0$.

\begin{remark}
Physically, time evolution is Markovian when it does not depend on the past
(or future) history\ of the quantum system and its environment -- there is
no \textquotedblleft memory\textquotedblright\ about the way it reaches its
present state. As far as I know, Markovian property holds for closed systems
and, as it was demonstrated by Davies in \cite{Da}, it holds also for open
systems under special conditions.
\end{remark}

If the system is Markovian and time reversal invariant, its quantum dynamical semigroup
can be extended to a one-parameter group in $\mathcal{S}\left( \mathcal{H}%
\right) $, with its group property being a consequence of the semigroup
property (\ref{def_quantum-dynamical-semigroup}):%
\begin{equation}
\tilde{\Gamma}_{t}:\mathcal{S}\left( \mathcal{H}\right) \rightarrow \mathcal{%
S}\left( \mathcal{H}\right) \ \ ,\ \tilde{\Gamma}_{t}:=\left\{ 
\begin{array}{r}
\Gamma _{t}\ \ \ \ ,\ t\geq 0 \\ 
\Gamma _{-t}^{-1}\ \ ,\ t<0%
\end{array}%
\right.  \label{def_quantum-dynamical-semigroup-extended}
\end{equation}

\medskip

\paragraph{Quantum dynamical semigroup's generator.}

\medskip

For technical reasons (to appear in subsection \ref{subsec:1.2}),
from now on assume the quantum dynamical semigroup can be extended to a%
\textit{\ semigroup of positive trace preserving superoperators\footnote{%
The term \textit{superoperator} is used for operators in a space of
operators.} in the space }$\mathcal{T}\left( \mathcal{H}\right) $\textit{\
of bounded trace class operators in }$\mathcal{H}$,%
\begin{equation}
\Gamma _{t}:\mathcal{T}\left( \mathcal{H}\right) \rightarrow \mathcal{T}%
\left( \mathcal{H}\right) \ \ ,\ t\geq 0
\label{def_time-evolution-map_extended}
\end{equation}%
Semigroup property:%
\begin{equation}
\Gamma _{t_{2}+t_{1}}=\Gamma _{t_{2}}\Gamma _{t_{1}}\ \ ,\ \forall
t_{2},t_{1}\geq 0  \label{def_time-evolution-map_semigroup}
\end{equation}%
Positivity:%
\begin{equation}
\Gamma _{t}\left( \sigma ^{\ast }\sigma \right) \geq 0\ \ ,\ \forall \sigma
\in \mathcal{T}\left( \mathcal{H}\right)
\label{def_time-evolution-map_positivity}
\end{equation}%
Trace preserving property:%
\begin{equation}
tr\left( \Gamma _{t}\left( \sigma \right) \right) =tr\left( \sigma \right) \
\ ,\ \forall \sigma \in \mathcal{T}\left( \mathcal{H}\right)
\label{def_time-evolution-map_trace}
\end{equation}%
Also, one assume the quantum dynamical semigroup is $\left\Vert .\right\Vert
_{1}$-continuous from above:%
\begin{equation}
\lim_{t\downarrow 0}\left\Vert \Gamma _{t}\left( \sigma \right) -\sigma
\right\Vert _{1}=0\ \ ,\ \forall \sigma \in \mathcal{T}\left( \mathcal{H}%
\right)  \label{def_time-evolution-map_continuity}
\end{equation}

The space $\mathcal{T}\left( \mathcal{H}\right) $ of trace class operators
in $\mathcal{H}$ is a Banach space w.r.t. the trace-norm $\left\Vert
.\right\Vert _{1}$, 
\begin{equation}
\left\Vert \sigma \right\Vert _{1}:=tr\sqrt{\sigma ^{\ast }\sigma }\ \ ,\
\forall \sigma \in \mathcal{T}\left( \mathcal{H}\right)
\label{def_trace-norm}
\end{equation}%
Moreover, $\mathcal{T}\left( \mathcal{H}\right) \subset \mathcal{B}\left( 
\mathcal{H}\right) $ (trace class operators are bounded) and the space of
finite-rank operators in $\mathcal{H}$ is $\left\Vert .\right\Vert _{1}$%
-dense in $\mathcal{T}\left( \mathcal{H}\right) $. For details, see \cite[%
pp.206-209]{RS1}.\footnote{%
Note that the notation of \cite{RS1} differs from our notation: in \cite{RS1}%
, $\mathcal{L}\left( \mathcal{H}\right) $ denotes the space of \textit{%
bounded operators} in $\mathcal{H}$, but here it denotes space of \textit{%
densely defined operators} in $\mathcal{H}$.}

\medskip

The above conditions are sufficient to guarantee that the quantum dynamical
semigroup has an infinitesimal generator:

\begin{theorem}[quantum dynamical semigroup's generator]
\label{teorema_generator}\ For the quantum dynamical semigroup (\ref%
{def_quantum-dynamical-semigroup}) under conditions (\ref%
{def_time-evolution-map_semigroup}), (\ref{def_time-evolution-map_positivity}%
), (\ref{def_time-evolution-map_trace}) and (\ref%
{def_time-evolution-map_continuity}), it holds:\newline
i) There exists an operator $L$ with $\left\Vert .\right\Vert _{1}$-dense
domain $Dom\left( L\right) \subset \mathcal{T}\left( \mathcal{H}\right) $
such that%
\begin{equation}
\lim_{t\downarrow 0}\left\Vert \frac{\Gamma _{t}\left( \sigma \right)
-\sigma }{t}-L\left( \sigma \right) \right\Vert _{1}=0\ \ ,\ \forall \sigma
\in Dom\left( L\right)  \label{def_quantum-dynamical-semigroup-generator}
\end{equation}%
ii)\ The quantum dynamical semigroup $\Gamma _{t\geq 0}$ is a limit of
exponentials related to $L$:%
\begin{equation}
\Gamma _{t}\left( \sigma \right) =\lim_{\varepsilon \downarrow 0}\left( \exp
\left( tL_{\varepsilon }\right) \sigma \right) \ \ ,\ \forall \sigma \in
Dom\left( L\right) ,\ \forall t\geq 0
\end{equation}%
where%
\begin{equation}
L_{\varepsilon }:=\frac{\Gamma _{\varepsilon }-id}{\varepsilon }\ \ ,\
\forall \varepsilon >0
\end{equation}%
iii)\ $L$ is bounded if and only if $Dom\left( L\right) =\mathcal{T}\left( 
\mathcal{H}\right) $. In this case, $\Gamma _{t\geq 0}$ is the exponential
of $L$:%
\begin{equation}
\Gamma _{t}\left( \sigma \right) =\exp \left( tL\right) \sigma \ \ ,\
\forall \sigma \in \mathcal{T}\left( \mathcal{H}\right) ,\ \forall t\geq 0
\end{equation}%
Further, in this situation it holds%
\begin{equation}
\lim_{t\downarrow 0}\left\Vert \Gamma _{t}-id\right\Vert _{1}=0
\label{equation_continuity}
\end{equation}%
In Quantum Mechanics, \textit{the generator} $L$\textit{\ }of the system's
quantum dynamical semigroup is called the system's \textit{Liouvillian
superoperator}.
\end{theorem}

For a proof, see \cite[pp.376-379]{Ru} or \cite[p.237-238]{Yo}. The limit (%
\ref{def_quantum-dynamical-semigroup-generator}) combined with the fact that
the operator norm and trace-norm satisfy \cite[p.209]{RS1}%
\begin{equation}
\left\Vert \sigma \right\Vert \leq \left\Vert \sigma \right\Vert _{1}\
,\forall \sigma \in \mathcal{T}\left( \mathcal{H}\right)
\end{equation}%
implies the \textit{system's equation of motion}, called \textit{Markovian
quantum master equation} \cite[p.119]{BP}:%
\begin{equation}
\frac{d}{dt}\rho \left( t\right) =L\left( \rho \left( t\right) \right)
\label{equation_markovian-master-equation}
\end{equation}

\begin{remark}
From the mathematical point of view, Theorem (\ref{teorema_generator})
generalizes \textit{Stone's Theorem}, which establishes the existence and
uniqueness of the infinitesimal generator for an one-parameter group of unitary
operators in Hilbert spaces -- for details see \cite[Theorem 13.38, p.382]%
{Ru}. From the physical point of view, Theorem (\ref{teorema_generator})
generalizes for completely positive Markovian quantum systems the von
Neumann equation for time-dependent states (\ref{equation_von-neumann}).
\end{remark}

\begin{remark}
The Hille-Yosida Theorem characterizes those operators in $\mathcal{H}$
which are generators of some dynamical semigroup \cite[Theorem 13.37, p.380]%
{Ru}, \cite[pp.246-249]{Yo}: a densely defined operator $L$ in a Banach
space $X$ is the generator of some trace-norm continuous one-parameter
semigroup of bounded operator in $X$ if and only if there are constants $C>0$
and $\gamma \in \mathbb{R}$ such that%
\begin{equation}
\left\Vert \left( \zeta I-L\right) ^{-m}\right\Vert _{\mathcal{B}\left(
X\right) }\leq C\left( \zeta -\gamma \right) ^{-m}\ \ ,\ \forall \zeta
>\gamma ,\ \forall m\in \mathbb{N}^{\mathbb{\ast }}
\label{equation_hille-yosida}
\end{equation}%
where $\left\Vert \ \right\Vert _{\mathcal{B}\left( X\right) }$ is the
operator norm in $\mathcal{B}\left( X\right) $.
\end{remark}

It is natural to ask about the general form of the Liouvillian of a
Markovian quantum system. Fortunately, this question has an answer for the
special class of quantum dynamical semigroups which are \textit{completely
positive}. This is the point of the following subsection.

\subsection{Completely positive quantum dynamical semigroups}
\label {subsec:1.2}

\textit{Completely positiveness} (see Appendix \ref%
{sec:a}) is a property of \textit{quantum
operations}, a concept of the theory of generalized measurements \cite[%
pp.85-89]{BP}, \cite{Kr1983}. This is a special property of a large class of
quantum systems, including subsystems of systems whose time evolution is
unitary and do satisfy some special conditions \cite[pp.122-123]{BP} \cite%
{Pe}. Such systems have a wide range of applications, ranging from quantum information theory \cite{TP} to astrophysics \cite{YZ} \cite{Yu} (to give only two examples).

If the Liouvillian of a completely positive quantum dynamical semigroup is
bounded, then it has a standard form, according to the following Theorems
due to Lindblad \cite{Li} (bounded Liouvillian) and also
Gorini-Kossakowski-Sudarshan \cite{GKS} (finite dimensional Hilbert space).
Although the finite dimensional case can be viewed as a special case of the
general Theorem due to Lindblad, it can be better scrutinized; so, I will
state them separately. As far as I know, those theorems have not been extended to the case of unbounded Liouvillians, a typical situation in physics.

\begin{theorem}[Standard form of a bounded Liouvillian \protect\cite{Li}]
\label{teorema_liouvillian-bounded} Let $L$ be the Liouvillian of the
quantum dynamical semigroup $\Gamma _{t\geq 0}$ of a completely positive
Markovian quantum system $\mathfrak{S}$ with Hilbert space $\mathcal{H}$.%
\newline
If $L$ is bounded (equivalently, $Dom\left( L\right) =\mathcal{T}\left( 
\mathcal{H}\right) $), then there exist a bounded self-adjoint operator $H$
in $\mathcal{H}$ and a countable family of bounded operators $%
V_{1},V_{2},... $ in $\mathcal{H}$ satisfying
\begin{equation}
\sum_{j}V_{j}^{\ast }V_{j}\in \mathcal{B}\left( \mathcal{H}\right)
\end{equation}%
such that%
\begin{equation}
L\left( \sigma \right) =-i\left[ H,\sigma \right] +\frac{1}{2}\sum_{j}\left(
V_{j}\sigma V_{j}^{\ast }-\frac{1}{2}\left( V_{j}^{\ast }V_{j}\sigma +\sigma
V_{j}^{\ast }V_{j}\right) \right) \ \ ,\ \forall \sigma \in \mathcal{T}%
\left( \mathcal{H}\right)  \label{form_liouvillian-infinite}
\end{equation}
\end{theorem}

\begin{theorem}[Standard Form of a Liouvillian in a finite dimensional Hilbert \protect\cite{GKS}]
\label{teorema_liouvillian-finite} Let $L$ be the Liouvillian of the quantum
dynamical semigroup $\Gamma _{t\geq 0}$ of a completely positive Markovian
quantum system $\mathfrak{S}$ with Hilbert space $\mathcal{H}$ having finite
dimension $N=\dim \mathcal{H}$. In this case $\mathcal{T}\left( \mathcal{H}%
\right) =\mathcal{B}\left( \mathcal{H}\right) =\mathcal{L}\left( \mathcal{H}%
\right) $ has dimension $N^{2}$.\newline
Let $\left( F_{j}\right) _{j=1...N^{2}}$ be a complete set in $\mathcal{B}%
\left( \mathcal{H}\right) $ with $F_{N^{2}}=I/\sqrt{N}$ which is orthonormal
w.r.t. trace, \textit{i.e.},%
\begin{equation}
tr\left( F_{i}^{\ast }F_{j}\right) =\delta _{ij}\ \ ,\ \forall
i,j=1,...,N^{2}
\end{equation}%
Then, there exist a self-adjoint operator $H$ in $\mathcal{H}$ and a unique
positive complex matrix $\left( a_{ij}\right) \in M_{N^{2}-1}\left( \mathbb{C%
}\right) $ such that%
\begin{equation}
L\left( \sigma \right) =-i\left[ H,\sigma \right] +%
\sum_{i,j=1}^{N^{2}-1}a_{ij}\left( F_{i}\sigma F_{j}^{\ast }-\frac{1}{2}%
\left( F_{j}^{\ast }F_{i}\sigma +\sigma F_{j}^{\ast }F_{i}\right) \right) \
\ ,\ \forall \sigma \in \mathcal{T}\left( \mathcal{H}\right)
\label{form_liouvillian-finite}
\end{equation}%
$\mathcal{\newline
}$Further, the matrix $\left( a_{ij}\right) $ is unique for each family $%
\left( F_{j}\right) $ and the operator $H$ is unique if it is required $%
tr\left( H\right) =0$.
\end{theorem}

\begin{remark}
The conditions on $\left( F_{j}\right) _{j=1}^{N^{2}}$ in Theorem \ref%
{teorema_liouvillian-finite} imply that all operators different from $%
F_{N^{2}}$ are traceless:\footnote{%
Proof: $\mathrm{tr}\left( F_{j}\right) =\mathrm{tr}\left( IF_{j}\right) =%
\sqrt{N}\mathrm{tr}\left( F_{N^{2}}^{\ast }F_{j}\right) =0\ \ ,\ \forall
j\in \left\{ 1,...,N^{2}-1\right\} $}%
\begin{equation*}
tr\left( F_{j}\right) =0\ \ ,\ \forall j=1,...,N^{2}-1
\end{equation*}
\end{remark}

\begin{remark}
Expression (\ref{form_liouvillian-finite}) of Theorem \ref%
{teorema_liouvillian-finite} can be transformed in the expression (\ref%
{form_liouvillian-infinite}) of Theorem \ref{teorema_liouvillian-bounded} by
a diagonalization of the positive-semidefinite matrix $\left( a_{ij}\right) $%
, as done in \cite{GKS} and \cite[pp.121-122]{BP}.
\end{remark}

\medskip

The operator $H$ is called \textit{Hamiltonian} and the term $-i\left[ H,.%
\right] $ is called the \textit{Hamiltonian part} of the system's
Liouvillian.The \textit{dissipator superoperator} is defined by (in
agreement with \cite[p.123]{BP}):%
\begin{equation*}
\mathcal{D}\left( \sigma \right) :=L\left( \sigma \right) +i\left[ H,\sigma %
\right]
\end{equation*}%
In the finite dimensional case,%
\begin{equation}
\mathcal{D}\left( \sigma \right) =\sum_{i,j=1}^{N^{2}-1}a_{ij}\left(
F_{i}\sigma F_{j}^{\ast }-\frac{1}{2}\left( F_{j}^{\ast }F_{i}\sigma +\sigma
F_{j}^{\ast }F_{i}\right) \right)
\label{def_dissipator-superoperator_finite-dimensional}
\end{equation}%
People call $\left( a_{ij}\right) $ the system's \textit{Kossakowski matrix} and
note that it has dimension $N^{2}-1$ when the Hilbert space has finite
dimension $N$.

\subsection{Hamiltonian's expectation value}
\label{subsec:1.3}
In this subsection I analyze Hamiltonian' expectation value values related
to time-dependent states for completely positive Markovian quantum systems.
Our considerations are restricted to the finite dimensional case, so $%
\mathcal{T}\left( \mathcal{H}\right) =\mathcal{B}\left( \mathcal{H}\right) =%
\mathcal{L}\left( \mathcal{H}\right) $ and the system's Liouvillian is given
by Lindblad's form (\ref{form_liouvillian-finite}).

\medskip

\begin{definition}[Dissipation Operator]
Consider a quantum system $\mathfrak{S}$ with Hamiltonian $H$ and dissipator
superoperator having the form (\ref%
{def_dissipator-superoperator_finite-dimensional}). The system's dissipation
operator is defined by the action of dissipator superoperator on the
Hamiltonian:%
\begin{equation}
D_{H}:=\mathcal{D}\left( H\right) =\sum_{i,j=1}^{N^{2}-1}a_{ij}\left(
F_{j}^{\ast }HF_{i}-\frac{1}{2}F_{j}^{\ast }F_{i}H-\frac{1}{2}HF_{j}^{\ast
}F_{i}\right)  \label{def_energy-dissipator-operator}
\end{equation}
\end{definition}

\begin{proposition}
\label{teorema_hamiltonian-time-derivative}\ Let $\rho \left( t\right) $ be
a time-dependent state of the system $\mathfrak{S}$. Then the related
time-dependent Hamiltonian's expectation value value satisfies the equation%
\begin{equation}
\frac{d}{dt}\left\langle H\mid \rho \left( t\right) \right\rangle =tr\left\{
\rho \left( t\right) D_{H}\right\}
\end{equation}

\begin{proof}
The Markovian quantum master equation (\ref%
{equation_markovian-master-equation}) with Liouvillian having the form (\ref%
{form_liouvillian-finite}) reads%
\begin{equation}
\frac{d}{dt}\rho \left( t\right) =-i\left[ H,\rho \left( t\right) \right] +%
\mathcal{D}\left( \rho \left( t\right) \right)
\end{equation}%
Since derivation commutes with trace, it holds%
\begin{equation}
\frac{d}{dt}\left\langle H\mid \rho \left( t\right) \right\rangle =\frac{d}{%
dt}tr\left\{ H\rho \left( t\right) \right\} =tr\left\{ H\frac{d}{dt}\rho
\left( t\right) \right\} =-itr\left\{ H\left[ H,\rho \left( t\right) \right]
\right\} +tr\left\{ H\mathcal{D}\left( \rho \left( t\right) \right) \right\}
\end{equation}%
Using that trace is invariant under cyclic permutation of factors, it follows%
\begin{equation}
tr\left\{ H\left[ H,\rho \left( t\right) \right] \right\} =tr\left\{ HH\rho
\left( t\right) -H\rho \left( t\right) H\right\} =tr\left\{ HH\rho \left(
t\right) -HH\rho \left( t\right) \right\} =tr\left\{ 0\right\} =0
\end{equation}%
Therefore%
\begin{equation}
\frac{d}{dt}\left\langle H\mid \rho \left( t\right) \right\rangle =tr\left\{
H\mathcal{D}\left( \rho \left( t\right) \right) \right\} =tr\left\{ \mathcal{%
D}\left( \rho \left( t\right) \right) H\right\}
\end{equation}%
Using (\ref{form_liouvillian-finite}) to expand the expression inside trace,
it follows%
\begin{eqnarray}
\frac{d}{dt}\left\langle H\mid \rho \left( t\right) \right\rangle
&=&tr\left\{ \mathcal{D}\left( \rho \left( t\right) \right) H\right\} \\
&=&\sum_{i,j=1}^{N^{2}-1}a_{ij}tr\left\{ F_{i}\rho \left( t\right)
F_{j}^{\ast }H-\frac{1}{2}F_{j}^{\ast }F_{i}\rho \left( t\right) H-\frac{1}{2%
}\rho \left( t\right) F_{j}^{\ast }F_{i}H\right\} \\
&=&\sum_{i,j=1}^{N^{2}-1}a_{ij}tr\left\{ \rho \left( t\right) F_{j}^{\ast
}HF_{i}-\frac{1}{2}\rho \left( t\right) HF_{j}^{\ast }F_{i}-\frac{1}{2}\rho
\left( t\right) F_{j}^{\ast }F_{i}H\right\} \\
&=&tr\left\{ \rho \left( t\right) \sum_{i,j=1}^{N^{2}-1}a_{ij}\left(
F_{j}^{\ast }HF_{i}-\frac{1}{2}F_{j}^{\ast }F_{i}H-\frac{1}{2}HF_{j}^{\ast
}F_{i}\right) \right\} \\
&=&tr\left\{ \rho \left( t\right) D_{H}\right\}
\end{eqnarray}
\end{proof}
\end{proposition}

\begin{corollary}
\label{teorema_disclosedness}\ The Hamiltonian's expectation value value is
constant for whatever be the system's time-dependent state if and only if
the system's dissipation operator is identically zero.

\begin{proof}
From Proposition \ref{teorema_hamiltonian-time-derivative}, the
Hamiltonian's expectation value value is constant whatever be the system's
time-dependent state if and only if $tr\left\{ \rho D_{H}\right\} =0\
\forall \rho \in \mathcal{S}\left( \mathcal{H}\right) $; this is equivalent
to $\left\langle D_{H}\phi \mid \phi \right\rangle =0\ \forall \phi \in 
\mathcal{H}$ and this is equivalent to $D_{H}=0$.
\end{proof}
\end{corollary}

The above result motivates the following definition:

\begin{definition}[Markovian Dispersive Quantum System]
A Markovian\ dispersive quantum system is a Markovian quantum system whose
dissipation operator is identically zero.
\end{definition}

\begin{remark}
In the finite dimensional case, Corollary (\ref{teorema_disclosedness})
characterizes dispersive quantum systems within the class of completely
positive Markovian quantum systems by a linear equation for its dissipation
matrix:%
\begin{equation}
\sum_{i,j=1}^{N^{2}-1}a_{ij}\left( F_{j}^{\ast }HF_{i}-\frac{1}{2}%
F_{j}^{\ast }F_{i}H-\frac{1}{2}HF_{j}^{\ast }F_{i}\right) =0
\label{equ_dispersiveness}
\end{equation}%
Just to emphasize, this equation is a necessary and sufficient condition for
the system's Hamiltonian to belong to the kernel of its dissipator
superoperator.
\end{remark}

In the next section, I present a simple example of Markovian dispersive
quantum system, showing that the class of such systems is non empty.

\section{Dispersive qubit}
\label{sec:2}

The theoretical discussion of previous section does not follow the usual
reasoning used in the modeling of physical systems. Instead, it just set the
mathematical framework for modeling of quantum systems. In general, one
associates to a quantum system its characteristic\textit{\ Hamiltonian} and,
in each specific situation that system is studied, its equation of motion is
constructed taking into account the interaction with other systems and all
relevant contributions to the system's dynamics due to its environment. This
procedure is used when we want to model the system's behavior under specific
conditions, or when we want to specify conditions for the system to behave
according to some prescription. For completely positive Markovian quantum
systems, one has to propose a Liouvillian and, to verify if that can be the
generator of the system's quantum dynamical semigroup, perform one of the
two following procedures:\newline
i)\ Check the Hille-Yosida Theorem's condition (\ref{equation_hille-yosida});%
\newline
ii)\ Solve the system's equation of motion (\ref%
{equation_markovian-master-equation}), built the dynamical semigroup and
compute its generator.

\medskip

Here I define \textit{dispersive qubit} as a two-level quantum system having
a special dynamics. The Hilbert space of this system is $\mathbb{C}^{2}$
with its usual structure of vector space and inner product. According with
the general framework, \textit{observables} are defined by \textit{%
self-adjoint operators} and \textit{states} are defined by \textit{density
operators} in $\mathbb{C}^{2}$. The system's equation of motion is the
Markovian master equation (\ref{equation_markovian-master-equation}) with
Liouvillian having form\ (\ref{form_liouvillian-finite}) in terms of a \textit{%
Hamiltonian} $H$ and a \textit{dissipator superoperator} $\mathcal{D}$:%
\begin{equation}
\frac{d}{dt}\rho \left( t\right) =-i\left[ H,\rho \left( t\right) \right] +%
\mathcal{D}\left( \rho \right)  \label{model_equation-motion}
\end{equation}%
The Hamiltonian has non-degenerated spectrum, with eigenvalues $E_{0} < E_{1}$. Using Dirac's notation, the corresponding normalized
eigenvectors of the Hamiltonian are written as $\left\vert E_{0}\right\rangle $ and $%
\left\vert E_{1}\right\rangle $; using the ordered basis $\left\{ \left\vert
E_{1}\right\rangle ,\left\vert E_{0}\right\rangle \right\} $, the space of linear
operators $\mathcal{B}\left( \mathbb{C}^{2}\right) $ is identified with the\
space of $2\times 2$ complex matrices $M_{2}\left( \mathbb{C}\right) $; in
particular, the Hamiltonian is given by the following diagonal matrix%
\begin{equation}
H=\left( 
\begin{array}{cc}
E_{1} & 0 \\ 
0 & E_{0}%
\end{array}%
\right)
\end{equation}%
I denote the identity matrix of $M_{2}\left( \mathbb{C}\right) $ and \textit{%
Pauli matrices} by%
\begin{equation}
\sigma _{0}:=\left( 
\begin{array}{cc}
1 & 0 \\ 
0 & 1%
\end{array}%
\right) \ \ ,\ \ \sigma _{1}=\left( 
\begin{array}{cc}
0 & 1 \\ 
1 & 0%
\end{array}%
\right) \ \ ,\ \ \sigma _{2}=\left( 
\begin{array}{cc}
0 & -i \\ 
i & 0%
\end{array}%
\right) \ \ ,\ \ \sigma _{3}=\left( 
\begin{array}{cc}
1 & 0 \\ 
0 & -1%
\end{array}%
\right)
\end{equation}%
Pauli matrices with identity matrix divided by $\sqrt{2}$ form a basis for $%
M_{2}\left( \mathbb{C}\right) $ which is orthonormal with respect to trace:%
\begin{equation}
tr\left\{ \frac{\sigma _{i}^{\ast }}{\sqrt{2}}\frac{\sigma _{j}}{\sqrt{2}}%
\right\} =\delta _{ij}\ \ ,\ \forall i,j=0,1,2,3
\end{equation}%
Therefore, one can use Pauli matrices to describe the system's Liouvillian
and the dissipator superoperator, as prescribed by Theorem \ref%
{teorema_liouvillian-finite} and defined by (\ref%
{def_dissipator-superoperator_finite-dimensional}):%
\begin{equation}
L\rho =-i\left[ H,\rho \right] +\mathcal{D}\left( \rho \right)
\end{equation}%
where%
\begin{equation}
H=\frac{E_{1}+E_{0}}{2}\sigma _{0}+\frac{E_{1}-E_{0}}{2}\sigma _{3}\ \ ,\ \ \mathcal{D}\left( \rho \right)
=\sum_{i,j=1}^{3}a_{ij}\left( \sigma _{i}\rho \sigma _{j}^{\ast }-\frac{1}{2}%
\left( \sigma _{j}^{\ast }\sigma _{i}\rho +\rho \sigma _{j}^{\ast }\sigma
_{i}\right) \right)
\end{equation}%
\label{model_hamiltoniano-dissipator}
and $\left( a_{ij}\right) $ is the system's Kossakowski matrix. Defining%
\begin{equation}
\Delta :=E_{1}-E_{0}  \label{model_energy-difference}
\end{equation}%
the system's \textit{dissipation operator} (\ref%
{def_energy-dissipator-operator}) is given by:%
\begin{equation}
D_{H}:=\frac{\Delta }{2}\sum_{i,j=1}^{3}a_{ij}\left( \sigma _{j}\sigma
_{3}\sigma _{i}-\frac{1}{2}\sigma _{j}\sigma _{i}\sigma _{3}-\frac{1}{2}%
\sigma _{3}\sigma _{j}\sigma _{i}\right) 
\label{model_dissipator-superoperator}
\end{equation}
Finally, to complete the definition of \textit{dispersive qubit system} I require that the dissipation operator is null:
\begin{equation}
D_{H} \equiv 0
\end{equation}%
This condition is an equation for Kossakowski matrix. Below, I deal with a special case.
\subsection{Special Dispersive Qubit}
\medskip
From now on, I will deal with the following special case of dispersive qubit.
For a fixed $\lambda >0$, called here \textit{dispersive parameter}, I define the Kossakowski matrix:
\begin{equation}
a_{ij}=\frac{1}{2}\lambda \delta _{i3}\delta _{j3}
\label{model_dissipation-matrix}
\end{equation}%
Finally, the explicit expression for the system's Liouvillian is%
\begin{equation}
L\rho =\frac{1}{2}\lambda \left( \sigma _{3}\rho \sigma _{3}-\rho \right) -i%
\frac{\Delta}{2}\left( \sigma _{3}\rho -\rho \sigma _{3}\right)
\label{model_liouvillian}
\end{equation}%
In matricial terms:%
\begin{equation}
\rho =\left( 
\begin{array}{cc}
\rho _{11} & \rho _{12} \\ 
\rho _{21} & \rho _{22}%
\end{array}%
\right) \leadsto L\rho =\left( 
\begin{array}{cc}
0 & -\left( \lambda +i\Delta\right) \rho _{12} \\ 
-\left( \lambda -i\Delta\right) \rho _{21} & 0%
\end{array}%
\right)  \label{model_liouvillian-matrix}
\end{equation}

\paragraph{Time evolution.}

\medskip

One can solve the equation of motion (\ref{model_equation-motion}) and
analyze the time evolution of states to get further details about what
happens to the system as time goes on. Writing the density matrix for a
generic time-dependent state%
\begin{equation}
\rho \left( t\right) =\left( 
\begin{array}{cc}
\rho _{11}\left( t\right) & \rho _{12}\left( t\right) \\ 
\rho _{21}\left( t\right) & \rho _{22}\left( t\right)%
\end{array}%
\right) \ \ ,\ t\geq 0
\end{equation}%
the system's equation of motion (\ref{model_equation-motion}) becomes

\begin{equation}
\left( 
\begin{array}{cc}
\dot{\rho}_{11}\left( t\right) & \dot{\rho}_{12}\left( t\right) \\ 
\dot{\rho}_{21}\left( t\right) & \dot{\rho}_{22}\left( t\right)%
\end{array}%
\right) =\left( 
\begin{array}{cc}
0 & -\left( \lambda +i\Delta\right) \rho _{12}\left( t\right) \\ 
-\left( \lambda -i\Delta\right) \rho _{21}\left( t\right) & 0%
\end{array}%
\right)
\end{equation}%
The solution of this equation is%
\begin{equation}
\rho \left( t\right) =\left( 
\begin{array}{cc}
\rho _{11}\left( 0\right) & e^{-\left( \lambda +i\Delta\right) t}\rho
_{12}\left( 0\right) \\ 
e^{-\left( \lambda -i\Delta\right) t}\rho _{21}\left( 0\right) & \rho
_{22}\left( 0\right)%
\end{array}%
\right) \ ,\ t\geq 0 \label{model_time-dependent-states_initial}
\end{equation}%
Under conditions which guarantee self-adjointness, positivity and trace one
for $2\times 2$ complex matrices, then the general form of the system's
time-dependent states is:%
\begin{equation}
\rho \left( t\right) =\left( 
\begin{array}{cc}
a & be^{-\left( \lambda +i\Delta\right) t} \\ 
\bar{b}e^{-\left( \lambda -i\Delta\right) t} & 1-a%
\end{array}%
\right) \ \ ,\ t\geq 0  \label{model_time-dependent-states}
\end{equation}%
where\footnote{%
We note that self-adjointness, positivity and trace one hold for all times $%
t\geq 0$ if and only if they hold for $t=0$.}%
\begin{equation}
a\in \mathbb{R},\ b\in \mathbb{C},\ 0\leq a\leq 1,\ a\left( 1-a\right) \geq
\left\vert b\right\vert ^{2}
\end{equation}

The system's quantum dynamical semigroup follows from (\ref%
{model_time-dependent-states}): 
\begin{equation}
\Gamma _{t}\rho =\left( 
\begin{array}{cc}
\rho _{11} & \rho _{12}e^{-\left( \lambda +i\Delta\right) t} \\ 
\rho _{21}e^{-\left( \lambda -i\Delta\right) t} & \rho _{22}%
\end{array}%
\right) \ \ ,\ \forall \rho =\left( 
\begin{array}{cc}
\rho _{11} & \rho _{12} \\ 
\rho _{21} & \rho _{22}%
\end{array}%
\right) \in \mathcal{L}\left( \mathcal{H}\right)
\end{equation}%
To verify the consistence of the model, we mention that $\Gamma _{t}$ is
actually a continuous semigroup (in the sense of (\ref%
{def_time-evolution-map_continuity})) and its generator is $L$:%
\begin{equation}
\lim_{t\downarrow 0}\frac{1}{t}\left( \Gamma _{t}\rho -\rho \right)
=\lim_{t\downarrow 0}\left( 
\begin{array}{cc}
0 & \rho _{12}\frac{e^{-\left( \lambda +i\Delta\right) t}-1}{t} \\ 
\rho _{21}\frac{e^{-\left( \lambda -i\Delta\right) t}-1}{t} & 0%
\end{array}%
\right) =L\rho \ \ ,\ \forall \rho \in \mathcal{L}\left( \mathcal{H}\right)
\end{equation}

\paragraph{Irreversibility.}

\medskip

The condition of positiveness will be violated in (\ref%
{model_time-dependent-states}) when $a<1$ and $\left\vert b \right\vert >0$ if
one extrapolates this solution for times before%
\begin{equation}
t_{\ast }:=\ln \frac{1-a }{\lambda \left\vert b \right\vert }
\end{equation}%
This fact suggests the model is non-time reversal invariant, since the quantum dynamical
semigroup cannot be naturally extended to a one parameter group! Actually,

\begin{proposition}
The dispersive qubit is non-time reversal invariant.

\begin{proof}
It is sufficient to show that there is no idempotent solution to the
time-reversing equation (\ref{def_reversible_time-reversing-equation}). So,
assume that equation (\ref{def_reversible_time-reversing-equation}) has some
solution $\Upsilon :\mathcal{T}\left( \mathcal{H}\right) \rightarrow 
\mathcal{T}\left( \mathcal{H}\right) $ and write%
\begin{equation}
\Upsilon \rho =\left( 
\begin{array}{cc}
\gamma _{11}\left( \rho \right) & \gamma _{21}\left( \rho \right) \\ 
\gamma _{12}\left( \rho \right) & \gamma _{22}\left( \rho \right)%
\end{array}%
\right) \ \ ,\ \forall \rho \in \mathcal{T}\left( \mathcal{H}\right)
\end{equation}%
For $\rho \in \mathcal{T}\left( \mathcal{H}\right) $, denote $\rho _{\ast
}:=\lim_{t\rightarrow \infty }\Gamma _{t}\rho $ as given by (\ref%
{model_stationary-state});\ then, from (\ref%
{def_reversible_time-reversing-equation}) it follows%
\begin{equation}
\Upsilon \rho =\lim_{t\rightarrow \infty }\left( \Gamma _{t}\Upsilon \Gamma
_{t}\rho \right) =\lim_{t\rightarrow \infty }\left( \Gamma _{t}\Upsilon
\lim_{t\rightarrow \infty }\Gamma _{t}\rho \right) =\left( 
\begin{array}{cc}
\gamma _{11}\left( \rho _{\ast }\right) & 0 \\ 
0 & \gamma _{22}\left( \rho _{\ast }\right)%
\end{array}%
\right) \ \ ,\ \forall \rho \in \mathcal{T}\left( \mathcal{H}\right)
\end{equation}%
Therefore $\gamma _{12}\equiv 0\equiv \gamma _{21}$ and this means $\Upsilon 
$ cannot be idempotent. This completes the proof.
\end{proof}
\end{proposition}

\begin{proposition}
The dispersive qubit has pure states which evolve to mixed states. More
precisely, an initial pure state remains pure if and only if it is one of
the Hamiltonian's eigenstates.

\begin{proof}
The density matrix for a pure state has the form%
\begin{equation}
\rho _{0}=\left( 
\begin{array}{cc}
\left\vert \lambda \right\vert ^{2} & \lambda \bar{\beta} \\ 
\bar{\lambda}\beta & \left\vert \beta \right\vert ^{2}%
\end{array}%
\right) \ \ ,\ \lambda ,\beta \in \mathbb{C},\ \left\vert \lambda \right\vert
^{2}+\left\vert \beta \right\vert ^{2}=1  \label{model_pure-state}
\end{equation}%
I note that $\det \rho _{0}=0$ and this condition is necessary for any
density matrix representing a pure state.%
\newline
The state $\rho _{0}$ is one of the two eigenstates of Hamiltonian if and
only if $\lambda =0$ or $\beta =0$; from (\ref{model_time-dependent-states}),
those states are pure and from (\ref{model_time-dependent-states}) remain
constant (and pure).\newline
However, any initially pure state $\rho \left( 0\right) =\rho _{0}$ with $%
\lambda \neq 0$ and $\beta \neq 0$ evolves to impure states, since%
\begin{equation}
\det \rho \left( t\right) =\det \left( 
\begin{array}{cc}
\left\vert \lambda \right\vert ^{2} & \lambda \bar{\beta}e^{-\left( \lambda
+i\Delta\right) t} \\ 
\bar{\lambda}\beta e^{-\left( \lambda -i\Delta\right) t} & \left\vert
\beta \right\vert ^{2}%
\end{array}%
\right) =\left\vert \lambda \right\vert ^{2}\left\vert \beta \right\vert
^{2}\left( 1-e^{-2\lambda t}\right) \neq 0\ \ ,\ \forall t>0
\end{equation}
\end{proof}
\end{proposition}

The system's stationary states are defined by the limit%
\begin{equation}
\lim_{t\rightarrow \infty }\Gamma _{t}\rho \ \ ,\ \rho \in \mathcal{S}\left( 
\mathcal{H}\right)
\end{equation}%
They are explicitly 
\begin{equation}
\rho _{a}:=\left( 
\begin{array}{cc}
a & 0 \\ 
0 & 1-a%
\end{array}%
\right) \ \ ,\ 0\leq a\leq 1  \label{model_stationary-state}
\end{equation}

The von Neumann entropy \cite[p.510]{CN} of the state (\ref%
{model_time-dependent-states}) is given by:%
\begin{eqnarray}
S\left[ \rho \left( t\right) \right] &=&-tr\left\{ \rho \left( t\right) \log
\rho \left( t\right) \right\} \\
&=&\log 2-a\log \left( 1+\sqrt{1-4\left( a\left( 1-a\right) -\left\vert
b\right\vert ^{2}e^{-2\lambda t}\right) }\right) + \\
&&-\left( 1-a\right) \log \left( 1-\sqrt{1-4\left( a\left( 1-a\right)
-\left\vert b\right\vert ^{2}e^{-2\lambda t}\right) }\right)
\end{eqnarray}%
As one can verify, $S\left[ \rho \left( t\right) \right] $ increases with
time when $b\neq 0$ and%
\begin{equation}
\lim_{t\rightarrow \infty }S\left[ \rho \left( t\right) \right] =\log
2-a\log \left( 1+\sqrt{1-4a\left( 1-a\right) }\right) -\left( 1-a\right)
\log \left( 1-\sqrt{1-4a\left( 1-a\right) }\right)
\end{equation}%
For the special case $a=1/2$, this limit reaches the entropy's maximum value 
\cite[p.513]{CN}:%
\begin{equation}
\lim_{t\rightarrow \infty }S\left[ \rho \left( t\right) \right] =\log 2
\end{equation}

\subsubsection{General observables}
\label{section-model-observables}

Let $X$ be an observable of the Dispersive Qubit, \textit{i.e.}, a
self-adjoint operator in $\mathbb{C}^{2}$. It has two eigenvalues (which can be equal) $%
x_{1}\geq x_{2}$ and corresponding orthonormal eigenvectors $\left\vert
x_{1}\right\rangle $ and $\left\vert x_{2}\right\rangle $:%
\begin{equation*}
X\left\vert x_{j}\right\rangle =x_{j}\left\vert x_{j}\right\rangle \ \ ,\ \
\left\langle x_{i}\mid x_{j}\right\rangle =\delta _{ij}\ \ ;\ i,j=1,2
\end{equation*}%
From above conditions and with an eventual redefinition of $\left\vert
x_{1}\right\rangle $ and $\left\vert x_{2}\right\rangle $, it follows that
there exists $\theta \in \left[ 0,\pi /2\right] $ such that%
\begin{equation}
\left\{ 
\begin{array}{l}
\left\vert x_{1}\right\rangle =\cos \theta \left\vert +\right\rangle +\sin
\theta \left\vert -\right\rangle  \\ 
\left\vert x_{2}\right\rangle =-\sin \theta \left\vert +\right\rangle +\cos
\theta \left\vert -\right\rangle 
\end{array}%
\right.   \label{model_mixing}
\end{equation}%
In the basis $\left\{ \left\vert E_{1}\right\rangle ,\left\vert E_{0}\right\rangle
\right\} $, we have%
\begin{equation*}
\left\vert x_{1}\right\rangle =\left( 
\begin{array}{c}
\cos \theta  \\ 
\sin \theta 
\end{array}%
\right) \ \ ,\ \ \left\vert x_{2}\right\rangle =\left( 
\begin{array}{c}
-\sin \theta  \\ 
\cos \theta 
\end{array}%
\right) 
\end{equation*}%
and%
\begin{equation*}
X=\left( 
\begin{array}{cc}
x_{1}\cos ^{2}\theta +x_{2}\sin ^{2}\theta  & \left( x_{1}-x_{2}\right) \sin
\theta \cos \theta  \\ 
\left( x_{1}-x_{2}\right) \sin \theta \cos \theta  & x_{1}\sin ^{2}\theta
+x_{2}\cos ^{2}\theta 
\end{array}%
\right) 
\end{equation*}

I remark that the observable $X$ is not compatible with the Hamiltonian $H$
when $\theta \neq 0$ and $\theta \neq \pi /2$, because they cannot be
simultaneously diagonalized.

\bigskip 

\paragraph{Time evolution of eigenstates of $X$.}

\medskip

If the initial state of the system is that corresponding to the eigenstate $%
\left\vert x_{1}\right\rangle $ of $X$%
\begin{equation*}
\rho _{x_{1}}\left( 0\right) =\left\vert x_{1}\right\rangle \left\langle
x_{1}\right\vert =\left( 
\begin{array}{cc}
\cos ^{2}\theta  & \cos \theta \sin \theta  \\ 
\cos \theta \sin \theta  & \sin ^{2}\theta 
\end{array}%
\right) 
\end{equation*}%
Then, according with (\ref{model_time-dependent-states_initial}) the
time-dependent state is given by%
\begin{equation*}
\rho _{x_{1}}\left( t\right) =\left( 
\begin{array}{cc}
\cos ^{2}\theta  & e^{-\left( \lambda +i\Delta\right) t}\cos \theta
\sin \theta  \\ 
e^{-\left( \lambda -i\Delta\right) t}\cos \theta \sin \theta  & \sin
^{2}\theta 
\end{array}%
\right) \ \ ,\ t\geq 0
\end{equation*}%
If $\theta \neq 0$ and $\theta \neq \pi /2$, than the state $\rho
_{x_{1}}\left( t\right) $ is a mixture for all $t>0$, since it cannot be put
in the form (\ref{model_pure-state}). In particular, the system evolves from
the pure state $\left\vert x_{1}\right\rangle \left\langle x_{1}\right\vert $
to the mixture%
\begin{equation*}
\lim_{t\rightarrow \infty }\rho _{x_{1}}\left( t\right) =\cos ^{2}\theta
\left\vert E_{1}\right\rangle \left\langle E_{1}\right\vert \ +\ \sin ^{2}\theta
\left\vert E_{0}\right\rangle \left\langle E_{0}\right\vert 
\end{equation*}

\paragraph{Expectation values of $X$.}

\medskip

The expectation value of $X$ w.r.t. a generic time-dependent state (\ref%
{model_time-dependent-states}) is 
\begin{equation}
\left\langle X\mid \rho \left( t\right) \right\rangle =\left. 
\begin{array}[t]{l}
a\left[ x_{1}\cos ^{2}\theta +x_{2}\sin ^{2}\theta \right] +be^{-\left(
\lambda +i\Delta\right) t}\left[ \left( x_{1}-x_{2}\right) \sin \theta
\cos \theta \right] + \\ 
+\bar{b}e^{-\left( \lambda -i\Delta\right) t}\left( x_{1}-x_{2}\right)
\sin \theta \cos \theta +\left( 1-a\right) \left[ x_{1}\sin ^{2}\theta
+x_{2}\cos ^{2}\theta \right] 
\end{array}%
\right.   \label{model_observable-expectation-value}
\end{equation}%
In particular, the stationary expectation value of $X$ is%
\begin{equation}
\lim_{t\rightarrow \infty }\left\langle X\mid \rho \left( t\right)
\right\rangle =\left\langle X\mid \lim_{t\rightarrow \infty }\rho \left(
t\right) \right\rangle =\left( a\cos ^{2}\theta +\left( 1-a\right) \sin
^{2}\theta \right) x_{1}+\left( a\sin ^{2}\theta +\left( 1-a\right) \cos
^{2}\theta \right) x_{2}
\label{model_observable-expectation-value_assymptotic}
\end{equation}%
Combining the above formulas, we get the expectation value of $X$ w.r.t. the
time-dependent state which started as the $X$'s vector-state $\left\vert
x_{1}\right\rangle $:%
\begin{equation}
\left\langle X\mid \rho _{x_{1}}\left( t\right) \right\rangle =\left( \cos
^{4}\theta +\sin ^{4}\theta \right) x_{1}+2\cos ^{2}\theta \sin ^{2}\theta
x_{2}+e^{-\lambda }\sin ^{2}\theta \cos ^{2}\theta \cos \left( i\Delta t\right) \left( x_{1}-x_{2}\right) 
\label{model_observable-expectation-value-time}
\end{equation}%
and%
\begin{equation}
\lim_{t\rightarrow \infty }\left\langle X\mid \rho \left( t\right)
\right\rangle =\left\langle X\mid \lim_{t\rightarrow \infty }\rho \left(
t\right) \right\rangle =\left( \cos ^{4}\theta +\sin ^{4}\theta \right)
x_{1}+2\cos ^{2}\theta \sin ^{2}\theta x_{2}
\label{model_observable-expectation-value-assimptotic}
\end{equation}

\paragraph{Probabilities of transition and of surviving.}

\medskip 

The transition probability from the $X$'s vector-state $\left\vert
x_{1}\right\rangle $ to the $X$'s vector-state $\left\vert
x_{2}\right\rangle $ after a time $t\geq 0$ is given by the expectation
value of the projection operator $\left\vert x_{2}\right\rangle \left\langle
x_{2}\right\vert $:%
\begin{equation*}
P\left( x_{1}\rightarrow x_{2};t\right) =tr\left( \left\vert
x_{2}\right\rangle \left\langle x_{2}\right\vert \rho _{x_{1}}\left(
t\right) \right) 
\end{equation*}%

Explicitly:
\begin{eqnarray*}
P\left( x_{1}\rightarrow x_{2};t\right)  &=&\left\langle E_{1}\mid
x_{2}\right\rangle \left\langle x_{2}\right\vert \rho _{x_{1}}\left(
t\right) \left\vert E_{1}\right\rangle +\left\langle E_{0}\mid x_{2}\right\rangle
\left\langle x_{2}\right\vert \rho _{x_{1}}\left( t\right) \left\vert
E_{0}\right\rangle  \\
&=&2\left[ 1-e^{-\lambda t}\cos \left( \Delta t\right) \right] \cos
^{2}\theta \sin ^{2}\theta 
\end{eqnarray*}

The surviving probability of the $X$'s vector-state $\left\vert
x_{1}\right\rangle $ after a time $t\geq 0$ is given by the expectation
value of the projection operator $\left\vert x_{1}\right\rangle \left\langle
x_{1}\right\vert $:%
\begin{equation*}
P\left( x_{1}\rightarrow x_{1};t\right) =tr\left( \left\vert
x_{1}\right\rangle \left\langle x_{1}\right\vert \rho _{x_{1}}\left(
t\right) \right) 
\end{equation*}%
Explicitly:
\begin{eqnarray*}
P\left( x_{1}\rightarrow x_{1};t\right)  &=&\left\langle E_{1}\mid
x_{1}\right\rangle \left\langle x_{1}\right\vert \rho _{x_{1}}\left(
t\right) \left\vert E_{1}\right\rangle +\left\langle E_{0}\mid x_{1}\right\rangle
\left\langle x_{1}\right\vert \rho _{x_{1}}\left( t\right) \left\vert
E_{0}\right\rangle  \\
&=&\cos ^{4}\theta +\sin ^{4}\theta +2e^{-\lambda t}\cos \left( \Delta t\right) \cos ^{2}\theta \sin ^{2}\theta 
\end{eqnarray*}%
One can easily verify that%
\begin{equation*}
0\leq P\left( x_{1}\rightarrow x_{2};t\right) ,P\left( x_{1}\rightarrow
x_{1};t\right) \leq 1\ \ ;\ \ P\left( x_{1}\rightarrow x_{2};t\right)
+P\left( x_{1}\rightarrow x_{1};t\right) =1  , \forall t\geq0
\end{equation*}

For later use, I rewrite above formulas:%
\begin{equation}
P\left( x_{1}\rightarrow x_{2};t\right) =\left[ \frac{1}{2}-e^{-\lambda
t}\left( \frac{1}{2}-\sin ^{2}\left( \frac{\Delta }{2}t\right) \right) %
\right] \sin ^{2}\left( 2\theta \right) 
\label{model_observable_transition-probability}
\end{equation}%
\begin{equation}
P\left( x_{1}\rightarrow x_{1};t\right) =1-\left[\frac{1}{2}-e^{-\lambda
t}\left( \frac{1}{2}-\sin ^{2}\left( \frac{\Delta }{2}t\right) \right) %
\right] \sin ^{2}\left( 2\theta \right) 
\label{model_observable_surviving-probability}
\end{equation}

I remark the role of the dispersive parameter $\lambda $ of dispersive qubit: it changes the initially \textit{time-dependent superposition} of the states of $X$ to a final (assymptotic) \textit{time-indepented mixture} of them!

\begin{figure}[t]
	\centering
		\includegraphics[width=0.30\textwidth]{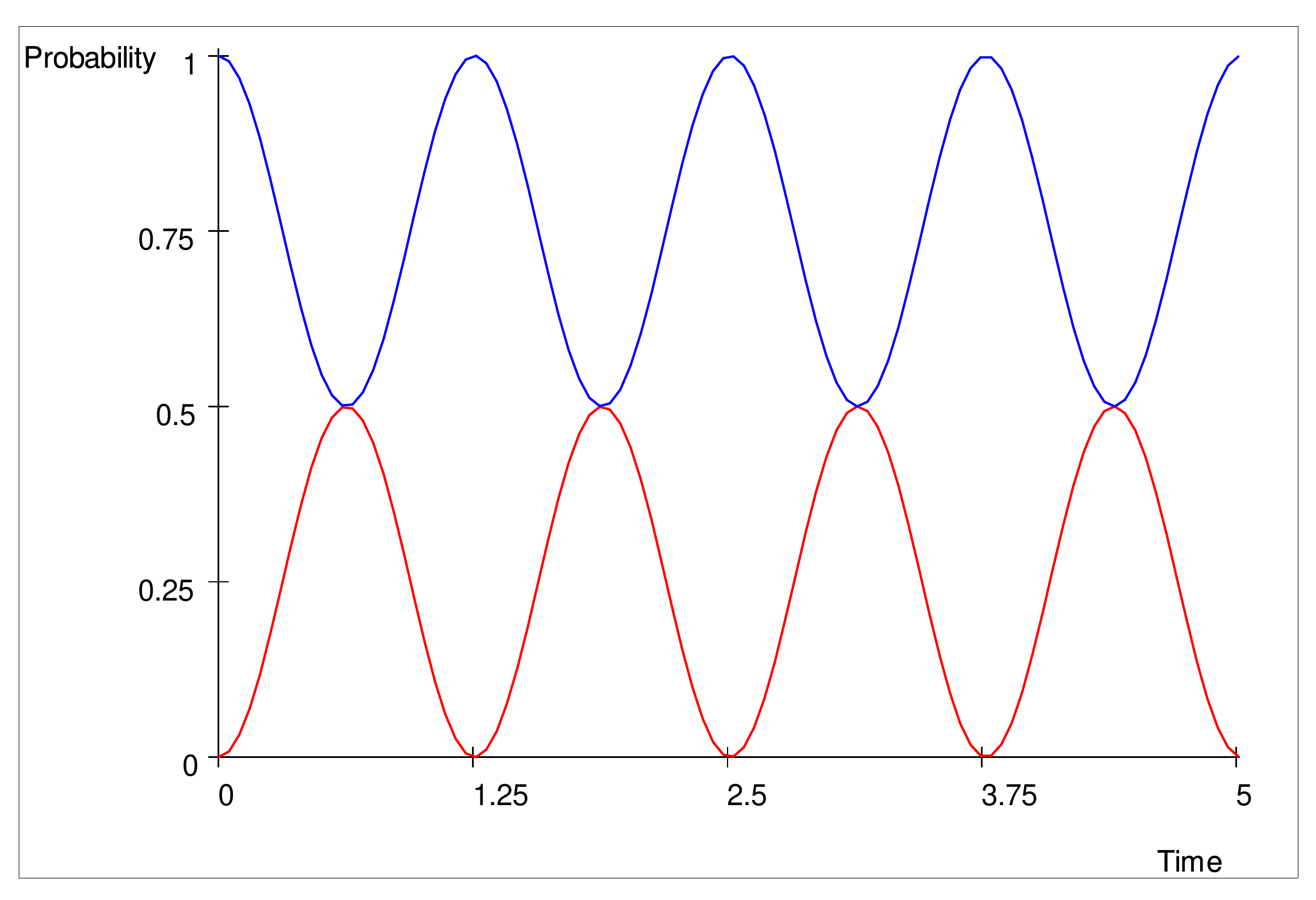}
	\caption{\small{Transition and surviving probabilities} for \small{$\Delta = 5 , \theta = \pi /8 , \lambda = 0$}}

	\label{fig:graph-0}
\end{figure}

\begin{figure}[t]
	\centering
		\includegraphics[width=0.30\textwidth]{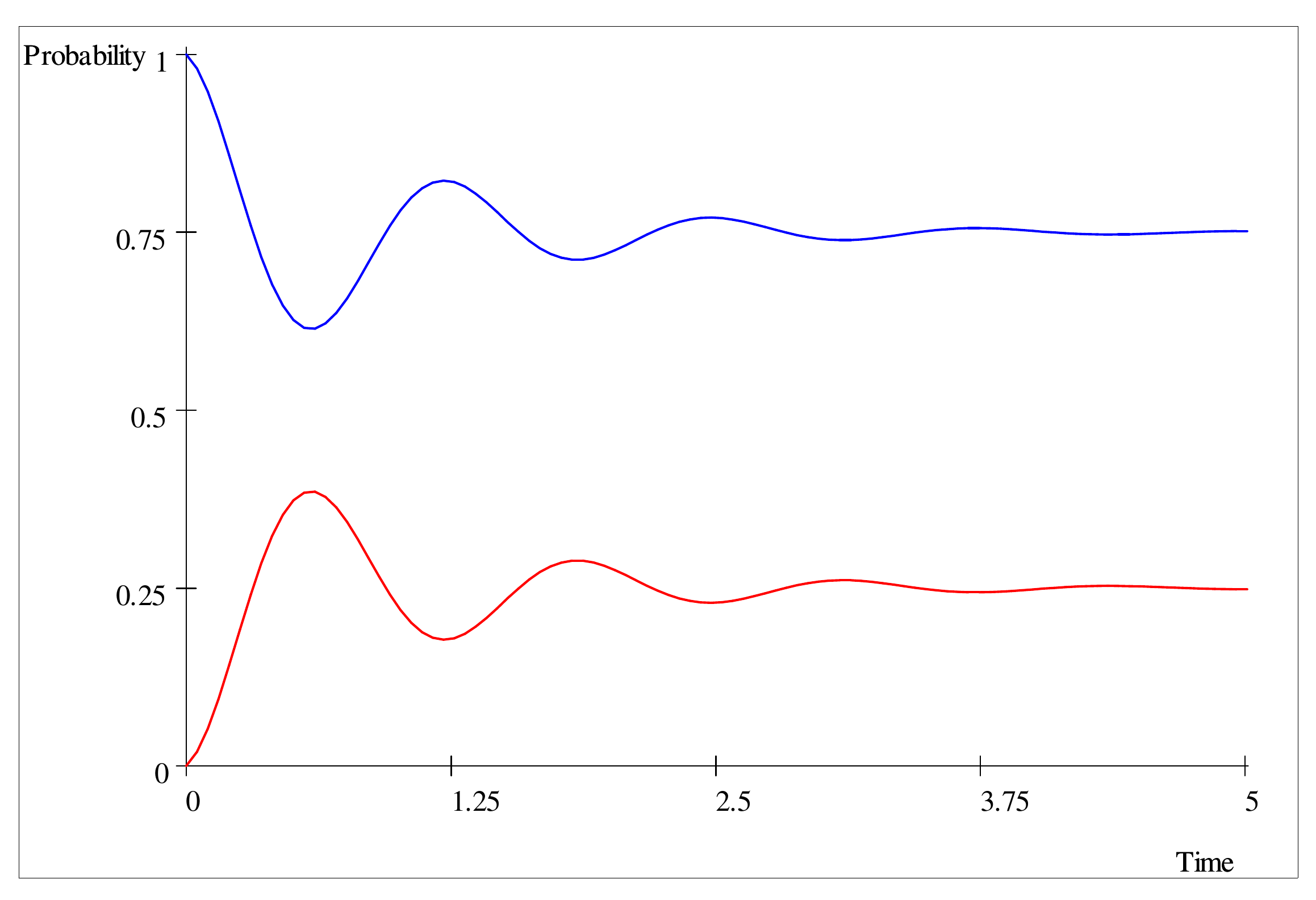}
	\caption{\small{Transition and surviving probabilities} for \small{$\Delta = 5 , \theta = \pi /8 , \lambda = 1$}}
	\label{fig:graph-1}
\end{figure}

\medskip
To ilustrate the time evolution of the special dispersive qubit, I plot in the fig.\ref{fig:graph-0} and fig.\ref{fig:graph-1} the graphs of the time-dependent transition and surviving probabilities corresponding to $\Delta=5 $ and $\theta =\pi /8$ for $\lambda =0$ and $\lambda =1$.

\medskip
\textit{Is there any application of the previous concepts and developments? Maybe...}

\subsubsection{Remark on the Kossakowski matrix}

\medskip

The following proposition shows that the choice (\ref%
{model_liouvillian-matrix}) for the Kossakowski matrix cannot be otherwise:

\begin{proposition}
The dispersive qubit's dissipator operator (\ref%
{model_dissipator-superoperator}) is zero if and only if its dissipation
matrix $\left( a_{ij}\right) $ satisfies 
\begin{equation}
a_{ij}=\beta \delta _{i3}\delta _{j3}\ \ ,\ \beta \geq 0
\end{equation}

\begin{proof}
We have to find all solutions to the equation for Kossakowski matrix's
coefficients%
\begin{equation}
\frac{1}{2}\Delta\sum_{i,j=1}^{3}a_{ij}\left( \sigma _{j}\sigma
_{3}\sigma _{i}-\frac{1}{2}\sigma _{j}\sigma _{i}\sigma _{3}-\frac{1}{2}%
\sigma _{3}\sigma _{j}\sigma _{i}\right) =0  \label{model_eq_dispersion}
\end{equation}%
subject to the conditions which guarantee self-adjointness%
\begin{equation}
\left( s.a\right) \ \bar{a}_{ij}=a_{ji}\ \ ,\ \forall i,j=1,2,3
\label{model_eq_sa}
\end{equation}%
and one of the following two sets of conditions which guarantee
positive-semi definiteness for $3\times 3$ matrices:%
\begin{equation}
\left. 
\begin{array}{l}
\left( i\right) :tr\left( a_{ij}\right) >0\ ,\ \det \left( a_{ij}\right) >0\
,\ \frac{1}{2}tr\left( a_{ij}\right) \left( \left( tr\left( a_{ij}\right)
\right) ^{2}-tr\left( \left( a_{ij}\right) ^{2}\right) \right) >\det \left(
a_{ij}\right)  \\ 
\left( ii\right) :tr\left( a_{ij}\right) >0\ ,\ \det \left( a_{ij}\right)
=0\ ,\ \frac{1}{2}tr\left( a_{ij}\right) \left( \left( tr\left(
a_{ij}\right) \right) ^{2}-tr\left( \left( a_{ij}\right) ^{2}\right) \right)
\geq 0%
\end{array}%
\right.   \label{model_eq_positivity}
\end{equation}%
\newline
For $\beta >0$, the matrix $\left( a_{ij}=\beta \delta _{i3}\delta
_{j3}\right) $ is positive since it obviously self-adjoint and has only
non-negative eigenvalues (namely, $0$ and $\beta $)\footnote{%
Equivalently, $\left( a_{ij}=\beta \delta _{i3}\delta _{j3}\right) $
satisfies conditions (\ref{model_eq_sa}) and (\ref{model_eq_positivity}-$ii$%
).}; by direct verification we see it satisfies equation (\ref%
{model_eq_dispersion}):%
\begin{equation}
\frac{\Delta \beta }{2}\left( \sigma _{3}\sigma _{3}\sigma _{3}-\frac{1}{2}\sigma
_{3}\sigma _{3}\sigma _{3}-\frac{1}{2}\sigma _{3}\sigma _{3}\sigma
_{3}\right)   \\
=  0
\end{equation}%
Now, it is very tedious to write down the calculations to get all solutions
for the above matrix equation (\ref{model_eq_dispersion}) subject to (\ref%
{model_eq_sa}) with one of the above two positivity conditions  (\ref{model_eq_positivity}); besides, such calculations are not directly relevant to the purposes of this paper; so I omit those here.
\end{proof}
\end{proposition}

\section{Dispersive model for neutrino oscillation}
\label{sec:3}
Neutrinos are neutral leptons which occur in one out of three flavors
(related to the others three charged leptons: electron, muon and tauon).
They have very tiny masses and interact extremely feebly, being sensible to
weak interaction and gravity only. Those characteristics mean that neutrinos
are \textquotedblleft quasi free\textquotedblright\ particles -- or more
precisely, they propagate almost unperturbed during the (eventually large)
time lapse between their production and detection. So, it is natural to pay
attention in neutrinos in the search to find a system combining both
\textquotedblleft isolatedness\textquotedblright\ and non-time reversal
invariance.

The phenomenology of neutrinos is not completely theoretically understood 
\cite{AS}. Actually, the Standard Model of Particle Physics and what must
extend/replace it are at stake \cite{Ho2007}. Specifically, the phenomenon
of \textit{neutrino oscillation} (defined as the dynamic change of flavor)
is a compelling evidence that those particles have masses different from
zero, an explicit contradiction with the Standard Model  \cite{Po}, \cite{MNS}, \cite{Fu},\cite{Kam}, \cite{PDG}. The theoretical mechanism explaining neutrino oscillation was
first devised by Pontecorvo \cite{Po} and \cite{MNS} around the 1960s; the
first experimental evidence\ of neutrino oscillation was obtained in the
Super-Kamiokande experiment in 1998 \cite{Fu}, and since them several
experiments have been realized around the world to measure the parameters
associated with neutrinos \cite{Kam}, \cite[pp.114-183]{PDG}, \cite{CNO}, 
\cite{Metal}.

Further, there are empirical evidences \cite{LSND2011} as well as
theoretical reasons \cite{PDG} for the existence of (at least) one more
neutrino flavor (besides the three standard flavors), called \textit{sterile
neutrino}. Presumably, sterile neutrino is a hight-handed particle which mix
itself with the other neutrino species and interacts only through gravity --
what makes it be a very ghostly particle. From the original proposal to
understand unexpected data concerning neutrino oscillation, some speculate
that sterile neutrino can also explain the large disparity of leptons masses
as well as be the reason behind the matter-antimatter asymmetry and,
further, that it can be the main missing ingredient of the Universe (as the
predominant component of dark matter) \cite[pp.114-183]{PDG}, \cite{Ji}. The
possibility of \textquotedblleft new physics beyond Standard
Model\textquotedblright\ is enlarged by unusual ideas to describe neutrino
dynamics and mixing of flavors: non-standard interactions \cite{Co2011} and
non-unitary time evolution of flavor-states \cite{An2006}, \cite{An2010}.

As long as there still are deep open questions about neutrinos, I think it
is opportune investigate the possibility that the dynamics of neutrinos can
be "dispersive", i.e., intrinsically non-time reversal invariant -- more
precisely: that the time evolution of neutrinos is non-time reversal
invariant even when they propagate isolately (in vacuum).

Specifically, I describe below the neutrino oscillation between two flavors%
\footnote{%
The oscillation between two neutrino flavors can be applied very well to the
solar neutrinos, since for them only the oscillation of $\nu _{e}$ and $\nu
_{\mu }$ is relevant} throughout a phenomenological approach similar to the
original ones \cite{Po} \cite{MNS}, called \textit{theory of massive and
mixed neutrino} in the monograph \cite{Bi} -- my basic reference.\footnote{%
Reference \cite{Li2006} also develops a more rigorous treatment which
justifies the probabilities of transition and surviving we find below (but
only in the \textit{dispersiveless} case, $\lambda =0$).} In this theory,
neutrino flavor states are superpositions of eigenstates of the
relativistic\ mass operator \cite[Chapter 4]{Bi}; in the quantum mechanical
approximation, the theory reduces to the case described in the subsection 
\textit{\ref{section-model-observables}}.

\bigskip

\paragraph{Dispersive theory of massive and mixed neutrino.}

\medskip

The oscillation between two neutrino flavors is described by a two-level
quantum system, identical to the \textit{dispersive quibit} defined in the
previous section. The flavor states, denoted by $\left\vert
\bar{\nu}_{e}\right\rangle $ and $\left\vert \bar{\nu}_{\mu }\right\rangle $, are
superpositions of the Hamiltonian's eigenvectors, denoted by $\left\vert
E_{{\nu}_{e}}\right\rangle $ and $\left\vert E_{{\nu}_{\mu }}\right\rangle $ through relation (%
\ref{model_mixing}), where $\theta $ is called the \textit{mixing angle} 
\cite[pp.107-108]{Bi}. Here\footnote{%
In the standard approach, one assumes that neutrino dynamics is given by Schr\"{o}dinger equation
with Hamiltonian $H$ \cite[p.99]{Bi}.}, I assume the dynamics is given by
Lindblad's equation (\ref{model_equation-motion}) with Hamiltonian and
dissipator superoperator (\ref{model_hamiltoniano-dissipator}) and
Kossakowski matrix (\ref{model_dissipation-matrix}) with a dispersive
parameter $\lambda $ (\ref{model_liouvillian}):%
\begin{equation*}
\frac{d\rho }{dt}=\frac{1}{2}\left( \lambda \sigma _{3}\rho \sigma _{3}-\rho
\right) -i\frac{\Delta }{2}\left( \sigma _{3}\rho -\rho \sigma _{3}\right)
\end{equation*}%
where (as in subsection \textit{\ref{section-model-observables}})%
\begin{equation*}
\Delta =E_{{\nu}_{\mu }}-E_{{\nu}_{e}}
\end{equation*}

Since neutrinos are produced with speed near the light velocity ($c=1$),
their energies must be given by the relativistic formula which combines mass
and momentum \cite[p.99]{Bi}: 
\begin{equation*}
E_{\chi}=\sqrt{p_{\chi}^{2}+m_{\chi}^{2}}\ \ ,\ \chi = {\nu}_{e},{\nu}_{\mu }
\end{equation*}

I assume the following conditions \cite[p.105]{Bi}:

i) The masses of neutrinos are small compared to the momenta they are
produced:%
\begin{equation*}
\frac{m_{ \chi }}{p_{ \chi }}\ll 1\ \ ,\ \chi = {\nu}_{e},{\nu}_{\mu }
\end{equation*}

ii)\ The momenta of neutrinos are approximately equal:%
\begin{equation*}
E:=p_{{\nu}_{e}}\simeq p_{{\nu}_{\mu }}
\end{equation*}

With above assumptions, it holds the approximations%
\begin{equation*}
E_{ \chi }\simeq p_{ \chi }+\frac{m_{ \chi }^{2}}{2p_{ \chi }^{2}}\simeq E+\frac{m_{ \chi }^{2}}{%
2E^{2}}\ \ ,\ \chi = {\nu}_{e},{\nu}_{\mu }
\end{equation*}%
and%
\begin{equation}
\Delta =E_{{\nu}_{\mu }}-E_{{\nu}_{e}}\simeq \frac{m_{{\nu}_{\mu }}^{2}-m_{{\nu}_{e }}^{2}}{2E}
\label{neutrino_energy-approx}
\end{equation}

The time lapse between the production and the detection of the neutrinos is
approximately given in terms of the distance $L$ between the source and the
detector by ($c=1$)%
\begin{equation}
t=L  \label{neutrino_time}
\end{equation}

Finally, substituting (\ref{neutrino_energy-approx}) and (\ref{neutrino_time}%
) in the formulas for the probability of transition (\ref%
{model_observable_transition-probability}) and surviving (\ref%
{model_observable_surviving-probability}) it follows with the insertion of
constants $c$ and $\hbar $:%
\begin{equation}
P\left( \bar{\nu}_{e}\rightarrow \bar{\nu}_{\mu };L,E,\lambda \right) =\left[ \frac{1}{2}%
-e^{-\lambda L/c}\left( \frac{1}{2}-\sin ^{2}\left( \frac{m_{1}^{2}-m_{0}^{2}%
}{4\hbar /c^{3}}\frac{L}{E}\right) \right) \right] \sin ^{2}\left( 2\theta
\right)  \label{neutrino_probability-transition}
\end{equation}%
and%
\begin{equation}
P\left( \bar{\nu}_{e}\rightarrow \bar{\nu}_{e};L,E,\lambda \right) =1-\left[ \frac{1}{2}%
-e^{-\lambda L/c}\left( \frac{1}{2}-\sin ^{2}\left( \frac{m_{1}^{2}-m_{0}^{2}%
}{4\hbar /c^{3}}\frac{L}{E}\right) \right) \right] \sin ^{2}\left( 2\theta
\right)  \label{neutrino_probability-surviving}
\end{equation}

In the case $\lambda =0$, these formulas are reduced to the standard ones \cite[pp.108-109]{Bi}.

Assuming $\lambda =0$, experimental data from KamLAND for the oscillation
between antineutrino-eletron $\bar{\nu}_{e}$ to the antineutrino-muon $\bar{\nu}_{\mu }$ gives
the following values for physical constants \cite{Kam}:%
\begin{equation*}
m_{{\nu}_{\mu }}^{2} - m_{{\nu}_{e }}^{2} = 7.9_{-0.5}^{+0.6} \times 10^{-5} \textit{eV}^{2} \ \ ,\ \  \tan ^{2} \theta = 0.40_{-0.7}^{+0.10}
\end{equation*}%
Finally, I leave to the experts the analysis of experimental data taking into
account the dispersive parameter:\footnote{Since the subject of this section is complex and out of my expertise, my intention is modest: illustrate the application of dispersive quantum system and incite further researches on the subject.} \emph{is it possible that the data can eventually corroborate the hypothesis that neutrino dynamics have a dispersive parameter different from zero?}

\section{Conclusions}
\label{sec:4}
To summing up, the effect of non-unitary time evolution in quantum systems can be threefold: \textit{%
dissipation/gain} (which means variation of system's energy), \textit{%
impurification} (which means time evolution from pure states to mixed
states, what is measured by the systems' entropy) and \textit{decoherence}
(which happens only to composed systems and was not discussed here). In general, those phenomena
are related to non-time reversal invariance and occur typically in open systems -- the environment being the
system's partner to the exchanges of matter, energy, momenta and information
(entropy). That is our expectation for completely positive Markovian quantum systems with non-zero dissipator superoperator (which means a deviation from unitary time evolution); however, for the class of dispersive quantum systems, time evolution implies impurification even when the systems are isolated.

\medskip

Surely, I don't know if the concept of dispersive quantum system can help us deepen our understanding about time reversal invariance and the meaning of \textit{irreversibility}. Nevertheless, the existence of an actual elementary dispersive quantum systems would be remarkable, since for them the non-time reversal invariance (and the property of the system's entropy be non-decreasing with time) cannot have a statistical meaning.

If one does not forget the essential difference concerning statistical interpretation, it can be useful to think about dispersive quantum systems as the quantum analogs of classical isolated thermodynamical systems (such as a low density gas in free expansion) because both can be isolated and non-time reversal invariant.

\medskip

I glimpse some developments to be done from what was presented here:\newline
i) The characterization of dispersive quantum systems with infinite degrees of freedom;\newline
ii) The study of decoherence in the context of dispersive quantum systems;\newline
iii) Improving the modeling of neutrino oscillation.\newline
iv) The building of new models and the search to get, in the laboratory, instances of dispersive quantum systems.\footnote{Perhaps, dispersive quantum systems can be manipulated in the laboratory by adjusting the environment into which a quantum system is inserted so that it behaves \textit{as if} it was isolated, up to a satisfactory degree.}

\medskip

Finally, the issue deserves further theoretical as well as experimental researches, if not due to the possibility to describe actual physical systems (like neutrinos), at least because we ignore answers to this simple question: \textit{If dispersive quantum systems cannot exist in nature (even approximately), are there physical principles forbidding them to exist?}

\appendix

\section{Completely positive quantum dynamical semigroups}
\label {sec:a}

Here, I present the definition of a \textit{completely positive quantum
dynamical semigroup} following \cite{GKS} (using a different notation). Denote by $M_{m}\left( 
\mathbb{C}\right) $ the C*-algebra of $m\times m$ complex matrices with
identity $I_{m}$, for any positive integer $m$.

\begin{definition}[Completely Positive Map]
\label{def_completely-positive-map} Let $\mathfrak{A}$ and $\mathfrak{B}$ be
C*-algebras. An operator $f:\mathfrak{A}\rightarrow \mathfrak{B}$ is said to
be completely positive if and only if for all positive integer $m$ the
following map is positive:%
\begin{equation}
f^{\left( m\right) }:=f\otimes I_{m}:\mathfrak{A}\otimes M\left( m\right)
\rightarrow \mathfrak{B}\otimes M\left( m\right)
\end{equation}
\end{definition}

\begin{definition}[Dual Map]
\label{def_dual-map} The dual of an operator $\Phi :\mathcal{T}\left( 
\mathcal{H}\right) \rightarrow \mathcal{T}\left( \mathcal{H}\right) $ is the
operator $\Phi ^{\ast }:\mathcal{B}\left( \mathcal{H}\right) \rightarrow 
\mathcal{B}\left( \mathcal{H}\right) $ defined by the following equation:%
\begin{equation}
tr\left( \sigma \Phi ^{\ast }\left( B\right) \right) =tr\left( B\Phi \left(
\sigma \right) \right) \ \ ,\ \forall \sigma \in \mathcal{T}\left( \mathcal{H%
}\right) ,\ \forall B\in \left( \mathcal{B}\left( \mathcal{H}\right) \right)
\end{equation}
\end{definition}

\begin{definition}[Dual Quantum Dynamical Semigroup]
The dual of the quantum dynamical semigroup $\Gamma _{t\geq 0}:\mathcal{T}%
\left( \mathcal{H}\right) \rightarrow \mathcal{T}\left( \mathcal{H}\right) $
is the one-parameter semigroup of dual maps: 
\begin{equation}
0\leq t\mapsto \Gamma _{t}^{\ast }:\mathcal{B}\left( \mathcal{H}\right)
\rightarrow \mathcal{B}\left( \mathcal{H}\right)
\end{equation}
\end{definition}

\begin{definition}[Completely Positive Quantum Dynamical Semigroup]
The quantum dynamical semigroup $\Gamma _{t\geq 0}$ is said to be completely
positive if and only if for all $t\geq 0$ the dual operator $\Gamma
_{t}^{\ast }$ is completely positive.
\end{definition}

\begin{remark}
In Quantum Mechanics, the dual quantum dynamical semigroup corresponds to the Heisenberg picture
for time evolution.
\end{remark}



\end{document}